\documentclass{article}
\usepackage[utf8]{inputenc}
\usepackage[letterpaper]{geometry}
\usepackage{float}
\usepackage{graphicx}
\makeatletter
\usepackage{gensymb}

\title{Directed aging, memory and Nature’s greed}
\author{Nidhi Pashine,\textit{$^{\dagger *a}$} Daniel Hexner,\textit{$^{* a,b}$} Andrea J. Liu,\textit{$^{b}$} Sidney R. Nagel\textit{$^{a}$}\\
\\
\normalsize{$^{a}$Department of Physics and The James Franck and Enrico Fermi Institutes,} \\
\normalsize{University of Chicago, Chicago IL, 60637}\\
\normalsize{$^{b}$Department of Physics and Astronomy,}\\
\normalsize{University of Pennsylvania, Philadelphia PA, 19104}\\
\normalsize{$^{*}$ Equal contribution. $^{\dagger}$ E-mail:  npashine@uchicago.edu.}}

\date{}
\makeatother

\begin{document}

\title{Directed aging, memory and Nature’s greed}
\maketitle

\begin{abstract}

Disordered materials are often out of equilibrium and evolve very slowly.  This allows a memory of the imposed strains or preparation conditions to be encoded in the material.  Here we consider ``directed aging'', where the elastic properties of a material evolve in the direction defined by its aging history.  The evolution to a lower-energy configuration is controlled by steepest decent and affects stressed regions differently from unstressed ones. This process can be considered to be a ``greedy algorithm" of Nature. Our experiments and simulations illustrate directed aging in examples in which the material's elasticity evolves as a direct consequence of the imposed deformation; the material itself decides how to evolve in order to produce responses that were not present inherently in the material. 
\end{abstract}

\section*{Introduction}

The incremental process of aging affects materials over extended periods
of time~\cite{Hodge_1995_Science,Struik_1977_Polymer,Hutchinson_1995_review,mitchell2008aging};  for example, a plastic may become
more brittle or a glass may slowly
shrink in volume. These changes occur without the apparent influence
of an external force directing the evolution. They are
simply part of the inexorable aging process, which often leads merely to a degradation
of physical properties. However, it comes as
no surprise that in other cases, such as a beam sagging under the
influence of gravity, an externally imposed stress forces the material
to deform over time in an obvious and well-defined manner. Here, while the geometry of the beam has perceptibly changed so that it may no longer do the job for which it was originally intended, changes in the material’s elastic properties are less apparent. 

Our purpose here is to highlight how the mechanics of materials can evolve
in unexpected ways due to a process of \textit{directed aging}. Materials that have not reached equilibrium often retain
a memory of how they were processed, trained, stored or manipulated~\cite{keim2018memory}. During aging, they evolve in a direction dictated by those conditions. The resulting properties contain a \textit{memory} of the applied load.
We unite these concepts
of directed aging and memory with the idea that Nature often follows
a \textit{greedy algorithm} in its evolution. 
We first recall how materials can be manipulated on the computer and then explore how natural evolution can be harnessed to produce a similar outcome. 

Materials can often be considered on an atomic scale as a network of bonds between nodes~\cite{Phillips, Thorpe}. Previous work has shown that networks can easily be tuned to have specific properties
by removing a small subset of  computationally determined links between nodes. 
For example, a spring network can be transformed from being nearly incompressible with
a positive Poisson’s ratio, $\nu \approx 0.5$, to being almost completely auxetic
with $\nu\approx-1$ (\textit{i.e.}, so that compression along one axis causes
the transverse directions to become equally compressed) \cite{CGoodrich_PRL_2015,DHex_PRE_2017,DHex_Softmatter_2018,DReid_PNAS_2018};
the network can be tuned over the entire spectrum of elastic
behavior into a regime where few materials exist.
Likewise, networks can be pruned for allosteric behavior so that a local applied strain
induces a large displacement at a distant point~\cite{JRocks_PNAS_2017,LYan_PNAS_2017,Tlusty_PRX_2017}.  

It takes surprisingly few alterations in the original network to create either of these functionalities.  
A take-away message is that structures generated from packings with rugged energy landscapes
are often extremely malleable -- with seemingly modest modifications, they can be easily manipulated to have unusual, esoteric and finely tuned properties.    

In these examples, a ``greedy''
computer algorithm was used at each stage to find the alteration (\textit{i.e.},
pruning of a bond) that brings the network closest to the desired
final state.
The algorithm does not require
-- nor necessarily benefit from -- a more sophisticated evolutionary
process that samples many alternative paths to reach an optimal outcome.  

Here we explore the extent to which a material in the process of aging can be considered as following
\textit{Nature’s} -- as distinct from a computer’s -- greediness to achieve
new properties. Can a material, without the intervention
of a computer, transform by retaining a distinct
memory of the forces it has
encountered in its lifetime?

\section*{Auxetic behavior derived from aging}
\subsection*{\textit{Gedanken} experiment}
We start with a highly idealized \textit{gedanken} experiment on a large heap of sand. Grains in this heap are under pressure from the material above it. Deep in the pile, the pressure is enormous and some of the contacts between grains experience immense forces. As they age, it is reasonable to expect that the contacts deform plastically, with those experiencing the largest forces deforming most rapidly. Over long times, these incremental deformations could become substantial and change the contacts between grains significantly.
While this seems straightforward, this system is interesting because there is \textit{preferential alteration} of the bond characteristics depending on the magnitude of stress that each individual contact feels under the applied stress. 

To gain insight into the effect of such preferential alteration, we recall the evolution under selective bond pruning of a spring network under compression. An idealized example of frictionless spherical grains jammed by compression~\cite{Liu_ARCMP_2010} can be converted into a network by replacing spheres with nodes and replacing contacts between spheres with unstretched springs connecting the nodes~\cite{alexander,Wyart_network}.
In such disordered networks, the contribution
of any specific bond to the bulk modulus, $B$, is to a large extent
independent of its contribution to the shear modulus, $G$~\cite{CGoodrich_PRL_2015}. If bonds are pruned according to how much stress they feel
due to an externally applied stress, it can alter the system’s bulk and shear
moduli in different ways~\footnote{We note that we prune a bond according to the stress it is under and not according to how much it would change a modulus
if it were removed. While the latter is a more effective way
to prune a network \cite{DHex_Softmatter_2018,DHex_PRE_2017}, it
is not essential.}. In particular, when placed under isotropic compression, pruning of 
bonds experiencing the largest stress causes $B$ to decrease
more rapidly than $G$ so that the ratio $G/B$ increases~\cite{CGoodrich_PRL_2015}. 

For an isotropic material, the Poisson’s ratio, $\nu$, is a monotonic
function of $(G/B)$: \\
$\nu=\frac{d-2(G/B)}{d(d-1)+2(G/B)}$ where $d$
is the spatial dimension. Therefore, as $(G/B)$ increases, the material
is driven to have a negative Poisson's ratio.
This result for pruned networks suggests that a sandpile under pressure could evolve towards auxetic behavior as well -- a surprising and novel outcome.  

Note that our \textit{gedanken} experiment  neglects any particle rearrangements that could occur during aging and is therefore valid only at the early stages before particle rearrangements occur.  The experiments and simulations we describe below also do not allow particle rearrangements.

\subsection*{Laboratory experiments}
The \textit{gedanken} experiment inspires the following experiments on aging under compression. 

\begin{figure}
\centering \includegraphics[width=150mm]{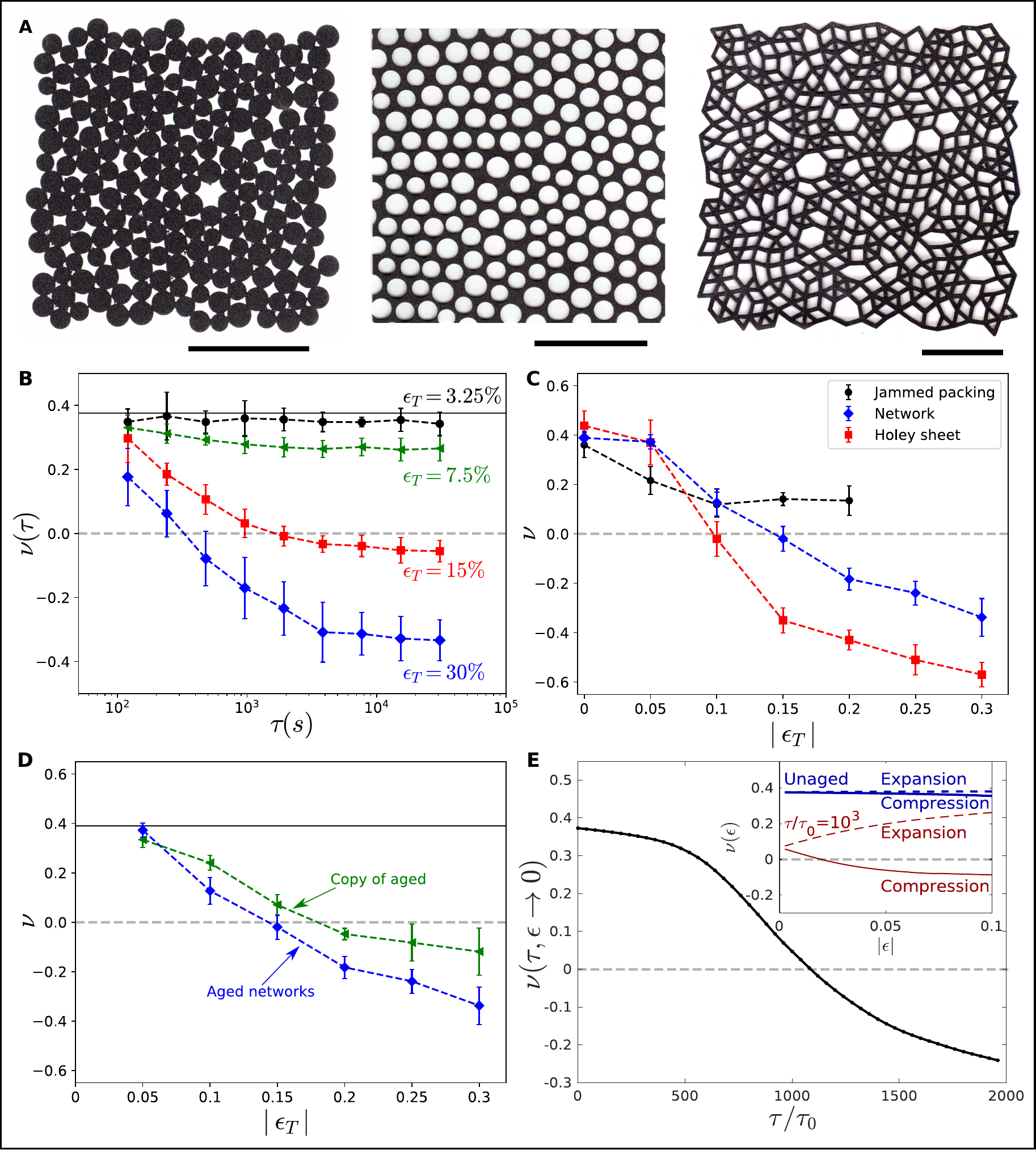} 

\caption{Change in Poisson's ratio, $\nu$, by aging. (A) Sample systems of each kind that were trained. Left to right: a jammed packing of discs, a network, and a disordered holey sheet. The scale bars are 50 mm.
(B) Effect of aging as a function of time (seconds). When foam networks are aged under constant strain, Poisson's ratio drops until it saturates to a final value.  Different curves represent different aging strains. The horizontal line at $0.38$ shows the initial $\nu$ of the networks. (C) Three different systems: jammed packings (black circles), networks (blue diamonds) and holey sheets (red squares) show a drop in $\nu$ when aged under uniform compression. Each data point is an average over multiple realizations and over the two perpendicular directions of applied strain. (D) Role of geometry: Networks prepared by aging under different strains (blue diamonds) compared to unaged networks cut out to have the same geometry (green squares). The deviation suggests that both geometry as well as the material properties change during aging.
(E) Results from numerical simulation of a system aged at 2.5\% compressive strain. The Poisson's ratio within linear response decreases as a function of time. Inset: The Poisson's ratio in the non-linear regime as a function of the measuring strain
for an unaged networks and networks aged at $\tau/\tau_0=10^3$ . Note that the system is not auxetic within linear response, but auxetic when compressed to larger strains.}

\label{network_nu} 
\end{figure}

From a sheet of EVA (ethylene vinyl acetate) foam, we laser-cut two-dimensional (2D) systems as shown in Fig.~\ref{network_nu}A.
We make three different kinds of systems: jammed packings of discs, networks derived from jammed packings, and random holey sheets, also derived from jammed packings.

In order to create the jammed packing, we start with a foam sheet and cut out a jammed configuration of discs obtained from a 2D computer simulation. The parts of discs that are overlapping with each other are left undisturbed, and this ensures that we have a fully connected sample. The networks are designed as explained in the previous section, where nodes represent particle centers and narrow struts connect nodes corresponding to overlapping pairs of particles. For the holey sheets, we again start with a jammed configuration, and shrink all the particles uniformly so they do not overlap. We cut holes corresponding to these smaller particles out of a sheet of EVA foam, leaving a sheet with a disordered pattern of holes. The jammed discs and holey sheets explicitly retain the circular nature of a 2D granular packing and are thus closer to the \textit{gedanken} sandpile experiment discussed above.

All three kinds of systems are then aged by confining them for a time $\tau$ in a square rigid box that has a smaller edge length, $L_{box}$, than the original length of the system, $L_{initial}$. We define the training strain as $\epsilon_{T} \equiv (L_{box} - L_{initial})/L_{initial}$. Since we train our samples under compression, our values for $\epsilon_{T}$ are always negative. We measure the Poisson's ratio, $\nu(\tau,\epsilon_{T})$, by removing our sample from the box, compressing it along one axis and measuring the deformation in the perpendicular direction. Fig.~\ref{network_nu}B shows data for Poisson's ratio of networks as a function of $\tau$. For $\mid\epsilon_{T}\mid \ge 0.15$, $\nu$ becomes negative at sufficiently long times. Fig.~\ref{network_nu}C shows how $\nu$ changes on aging for sufficiently long times. All three systems show that $\nu$ decreases as a function of $\mid \epsilon_T\mid$, as predicted for our \textit{gedanken} experiment. For networks and holey sheets, $\nu$ becomes negative at larger values of $\mid\epsilon_{T}\mid$.

In this example of directed aging, the material naturally acquires an auxetic response. 
The fact that all three systems show similar behavior suggests that directed aging of isotropic disordered systems under compression may lead more generally to reduced values of the Poisson's ratio.

There are two ways in which aging may have affected the network: (1) bond strengths change due to the stresses to which they were exposed, and (2) the geometry of the network changes due to internal rotation, deformation and/or buckling of bonds. To assess the relative contributions of these two effects, we image an \textit{aged} network. We use that image to pattern
another network out of \textit{unaged} material that is as geometrically identical as possible to the aged one. The difference between the Poisson's ratio of these
two (nearly) identical networks, is presumably due solely to
the aging of the stiffness of the contacts. 

The Poisson's ratio of these geometrically identical but unaged networks is shown in green triangles in Fig.~\ref{network_nu}D.
We consistently find that the \textit{unaged, copied} networks have a lower
Poisson's ratio than the original networks but not as low as that of the aged networks. This implies that some of the contribution
to the aging process derives from a change in the network geometry and some is due to a change in the material stiffness.

Our experimental protocol is similar to earlier work by Lakes~\cite{RLakes_Science_1987}, in which foam was made auxetic by heating it while under compression. In that work, the geometry of the structure evolved and concave polygons were observed after the foam was returned to ambient temperature. 
The evolution of auxetic behavior was attributed to the creation of concave polygons, which are known to decrease the Poisson's ratio of a material~\cite{Gibson_1982,DReid_PNAS_2018,Lakes_2017}.  In contrast, our results show that geometrical changes are not the only contribution; changes in bond stiffness also participate in creating auxetic behavior.

\subsection*{Aging in simulations}

We consider a network in which each spring, $i$, has spring constant $k_{i}$ and an unstretched length $l_{i}^{0}$. When compressed, the spring has length $l_{i}$. Due to disorder, each spring will in general be compressed by a different amount.
The energy of the resulting network is the sum of the energies of all the springs:

\begin{equation}
E=\frac{1}{2}\sum_{i}k_{i}(l_{i}-l_{i}^{0})^{2}\label{eqn:energy}
\end{equation}
For simplicity, we have omitted the energy due to bending  angles around a node~\cite{DReid_PNAS_2018}.   

We consider two limiting cases. The first corresponds to the case in which aging is due completely to evolution of the bond strengths $k_i$ under the imposed stresses.  The second corresponds to aging arising completely from the evolution of the equilibrium distances between nodes, $\ell_{i}^{0}$, as proposed by Lakes \cite{RLakes_Science_1987}. 

Here we focus on the first case, deferring the second to a future publication. 
We evolve $k_i$ so that it decreases in time according
to the elastic energy stored in that bond:

\begin{equation}
\frac{dk_{i}}{d\tau}=-\frac{1}{\tau_0 \bar{l_i^0}^2}k_{i}(l_{i}-l_{i}^{0})^{2}\label{eqn:dk_i}
\end{equation}
The proportionality constant, $\tau_0$ and the average equilibrium length of the bonds, $\bar{l_i^0}$, are material dependent and set the units of time and length. Bonds under greater stress evolve more rapidly. At late times, $k_i \rightarrow 0$; we therefore consider Eq.~\ref{eqn:dk_i} as an approximation valid only at early times.  
We evolve the system at a prescribed training strain, $\epsilon_{T}$,
and then measure the elastic response with respect to the zero-strain state, which remains the global energy minimum. 

Fig.~\ref{network_nu}E shows the
evolution of the Poisson's ratio $\nu$ calculated at $\mid\epsilon_T\mid \rightarrow 0 $, when the system is aged under compression. Evidently, $\nu$
decreases in time, consistent with the experimental data in Fig.~\ref{network_nu}C, and eventually becomes negative. The aging evolves in a directed manner -- under compression the bulk modulus $B$ decreases with respect to the shear modulus, $G$, leading to an auxetic material. 

This model also predicts \textit{non-linear} behavior. The nonlinear Poisson's ratio is defined by $\nu\left(\epsilon\right)=-\epsilon_{r}/\epsilon$,
where $\epsilon$ is the imposed strain along one axis
and $\epsilon_{r}$ is the resulting transverse strain after minimizing the energy with respect to %the box shape and 
the locations of all the nodes and the box shape. The inset of Fig.~\ref{network_nu}E shows $\nu(\epsilon)$ for the unaged
network as well as for a network that has been aged until $\tau/\tau_0=10^3$. The original, unaged network depends only weakly on the measuring strain $\epsilon$, even up to 10\% strain.
For aged networks, however, the
nonlinear Poisson's ratio, $\nu\left(\epsilon\right)$ depends  on $\epsilon$, the strain at
which it is measured, as shown in Fig.~\ref{network_nu}E.  The material can become  auxetic \textit{even when it is not auxetic in linear response.} In other words, compressing the system along one axis leads to transverse \textit{expansion} at small strains, and to transverse \textit{compression} under larger strains. This nonlinear behavior is difficult to achieve by design, but here occurs naturally. 

Both the experiments and simulations demonstrate that aging
is \textit{directed} towards a non-trivial elastic state with a lower  Poisson’s
ratio. 
The elastic
properties start to change as soon as aging commences under the applied
stress and evolve to be significantly different from those of
the freshly-prepared material.

\section*{Directed aging under shear}

Not all aging protocols evolve towards the same limit. Here, we evolve a system under shear stress instead of compression.
We return to the experimental networks described in Sec.~2.2, aging our networks under pure shear by compressing them in one direction
and stretching them in the perpendicular direction.

We measure the
Poisson's ratio in two perpendicular directions: (i) compressing the network
first along one axis and measuring the response along the second axis and (ii) exchanging which axis is compressed and which measured. 

\begin{figure}
\centering \includegraphics[width=100mm]{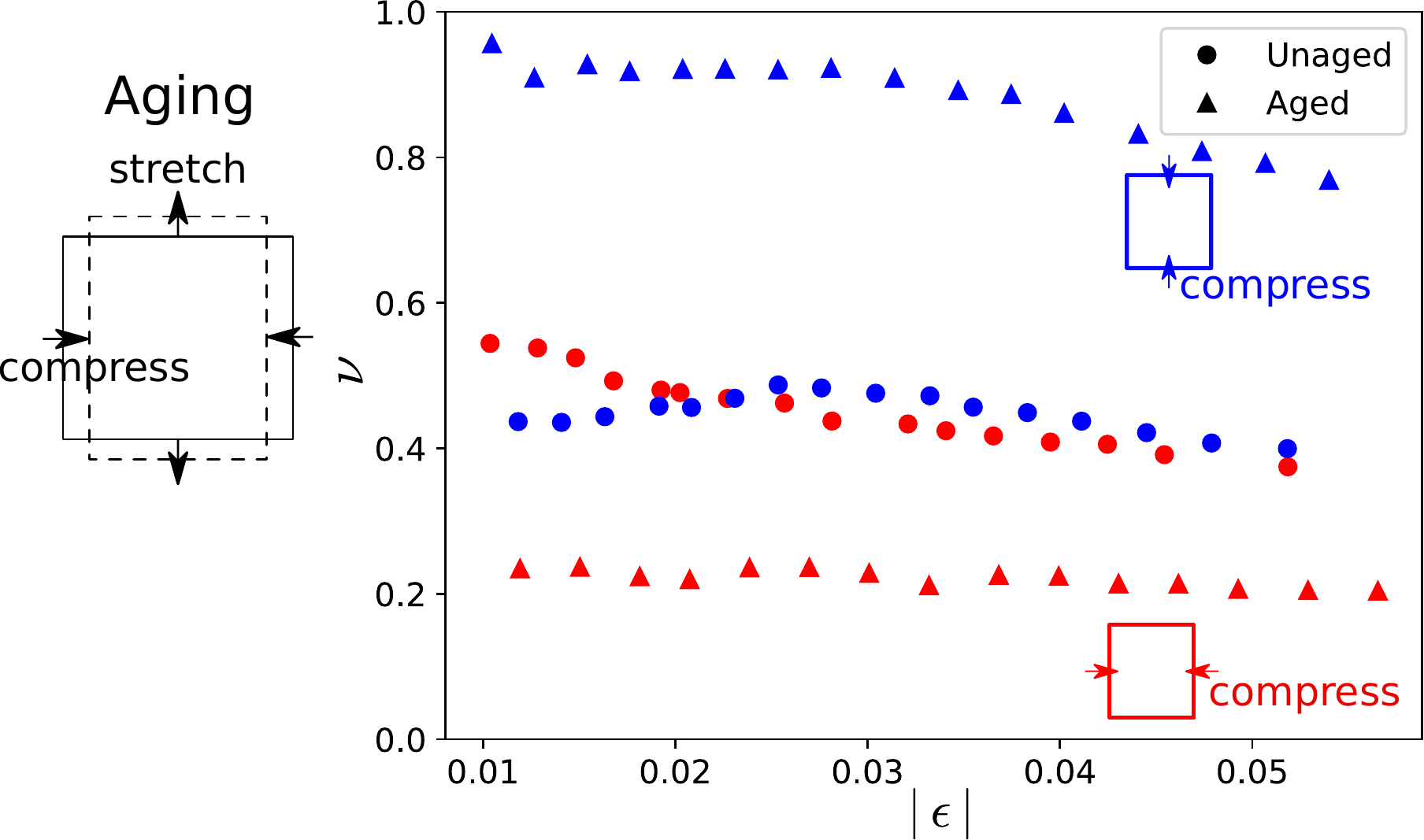} \caption{The Poisson's ratio of experimental foam networks aged under pure shear strain.  The abscissa is the strain at which Poisson's ratio is measured when the network is compressed along a given direction. The unaged (circles) networks do not show a significant dependence on measurement strain along either axis. 
The aged networks (triangles) show a dramatic change in behavior when $\nu$ is measured by compressing along the pulling axis (blue) or along the axis that was compressed (red) during aging.}
\label{shear} 
\end{figure}

Initially, the Poisson's ratio of the network measured by compressing
the network along either direction gives $\nu\approx0.4$. However, once it has been aged under shear, the same
measurements give very different results. The Poisson's ratio drops
to $\nu\approx 0.2$ when compressed in the same direction as the one along which
it was aged. However, when compressed along the perpendicular direction, $\nu$ \textit{increases} from $\nu \approx 0.4$ to $\nu \approx 0.8-0.9$. Fig.~\ref{shear} shows that the material encodes a memory of the direction in which it was aged. Similar anisotropic behavior has been observed in a crosslinked actin network under shear~\cite{SMajumdar2018mechanical}. 

\section*{Response under local distortion}

So far, we have considered a material's response to the global distortions of compression and pure shear. Here we show that directed aging can train a more exotic response by applying local as well as global strains during aging.

\begin{figure}
\centering \includegraphics[width=55mm]{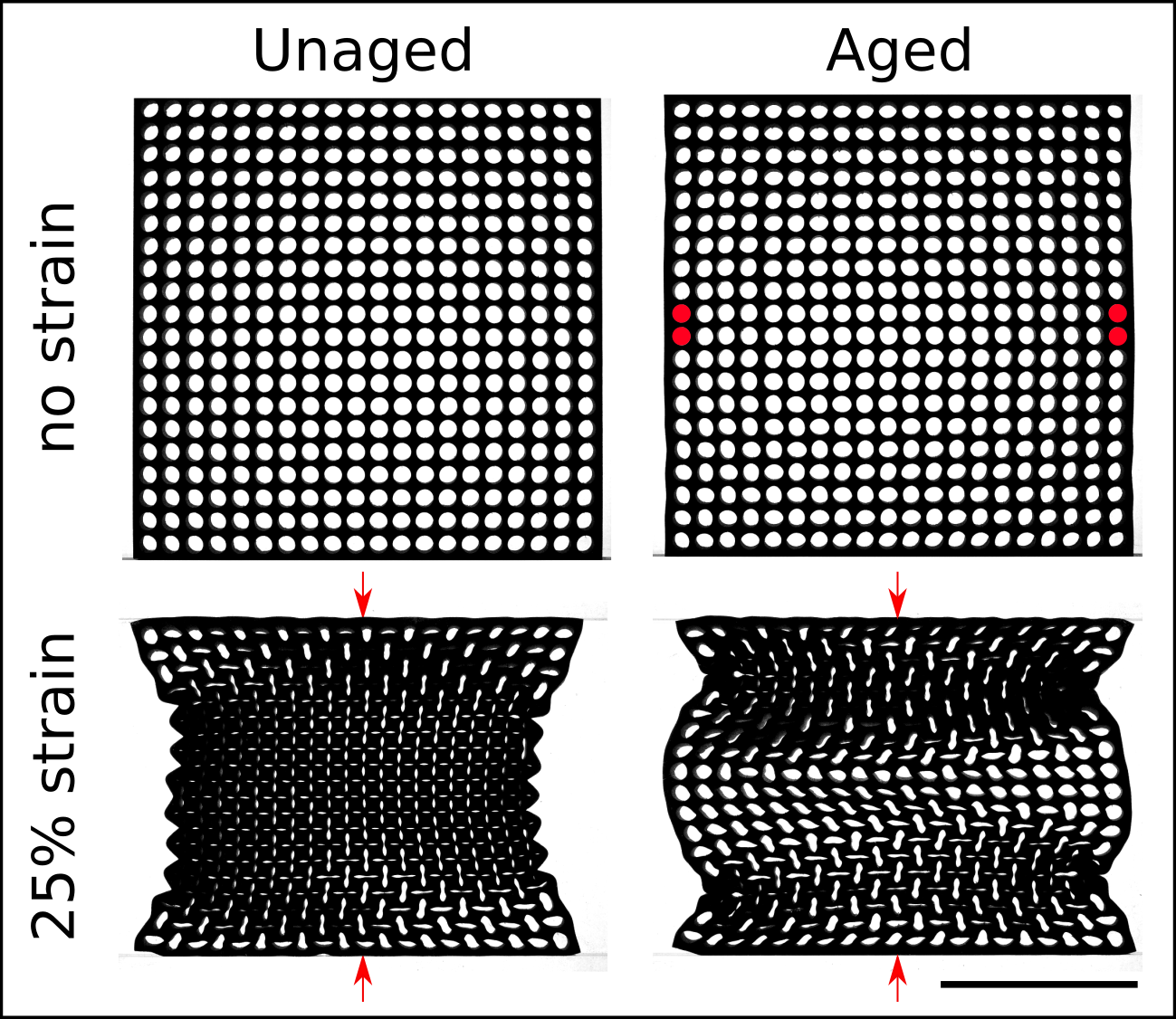} \caption{A holey sheet with an array of holes in a square lattice pattern. Initially the sheet is auxetic under compression along one of the major axes. If the sheet is aged while fixing 4 holes (shown in red)
while the sheet is under compression, the sheet now responds in a scalloped pattern when compressed in the vertical direction. The scale bar is 100mm. 
}
\label{holey_figure} 
\end{figure}

Sheets with a square lattice of circular holes~\cite{TMullin_PRL_2007,JShim_SoftMatter_2013,BFlorijn_PRL_2014}
have $\nu<0$ when compressed along the axis directions
\cite{TMullin_PRL_2007}. When this ``holey sheet'' is compressed, alternating holes distort in perpendicular
directions. This ordered collapse results in the transverse
edges of the sheet moving closer together, making the sheet
auxetic as shown in Fig.~\ref{holey_figure}A.

To obtain a more varied response, we age the sheet in the following way.  We plug the holes at the
end of the middle two rows so that they retain their circular shape
and stay in place as we compress the system along the direction perpendicular
to these rows. These rows separate the sheet into two halves that
do not communicate with each other. Approximately half of the time that these
sheets are compressed the pattern of holes in one  half is out of phase with those in the
other so that
the central region does not buckle inward.  In these cases,
compression along the perpendicular direction creates an overall scallop
pattern along the side edges where the sheet juts out in the middle as shown in Fig.~\ref{holey_figure}. 

\section*{Directed aging during cyclic loading}

Directed aging occurs in a host of other contexts. For example, it has been shown that cyclic shearing of a suspension of non-Brownian particles can drive the system towards a ground state where, under subsequent shearing, it simply repeats its former motion \cite{DPine_Nature_2005,LCorte_NatureP_2008,NKeim_PRL_2011,NKiem_PRE_2013,JPaunlsen_PRL_2014}.  This can also be seen in glassy and jammed solids~\cite{NKiem_Softmatter_2015,DFiocco_PRL_2014,HNagamanasa_PRE_2014,Sood_arXiv_2018,MLavrentovich_PRE_2017}. Cyclic driving produces a memory of how the system was prepared. As the cyclic shearing is applied, the training cycles force rearrangements in the particle positions.

The system ages in a directed fashion in that it eventually learns how to navigate the phase space to produce a periodic response. In some cases, it can find the periodic orbit rapidly; in other cases, it takes many training cycles before it finds an orbit with possibly higher-order periodicity~\cite{royer2015precisely,corey2008reversible,HNagamanasa_PRE_2014,MLavrentovich_PRE_2017}. Nevertheless, the system directly ages to develop this highly unusual dynamic property. It does so by a locally greedy algorithm of nature -- the system simply minimizes its energy (or enthalpy) at each time step (in the case of the solid) or it simply pushes other particles out of the way (in the case of the suspensions). The way in which the material is handled directs the evolution into non-generic states.

\section*{Conclusion}

We have demonstrated that aging, which is typically considered
to be a detrimental process, can be harnessed to encode a desired elastic
response in a material. This directed-aging process relies on incremental changes in stiffness and distortion of the microscopic structure brought about by plastic deformation.
We have given several examples in which directed aging exploits the structural disorder of the initial state to achieve a broad range of responses.  
Using these ingredients, we can then tune the linear and  \textit{non-linear} response simply by controlling the boundary conditions during the aging process.  

The idea of directed aging can be more general, as shown by the scalloped pattern that emerges from the compression of the sheet with a lattice of circular holes.  Simply letting a system age under specially-constructed constraining boundary conditions can encode novel functions.

This suggests that material properties, which are normally considered to be a function of the material composition and structure, can depend strongly on the history of the imposed deformations.  
One could ask whether this dependence on history could be used to read out the history of a material. For example, could we learn from the elasticity of a rock about geological flows that occurred over millions of years?

The different elastic responses we have discussed have been trained with remarkable ease. Designing elastic properties numerically typically requires a detailed knowledge of the precise interactions, as well as intensive optimization of the parameters. Here
we showed that despite the complex structure of the networks, and the
nonlinear strains at which they were aged, all that is required is patience while the system memorizes the training conditions and ages towards the target state.
This makes our approach well suited to designing materials with novel
functionality and elastic properties~\cite{DReid_PNAS_2018,JRocks_PNAS_2017,LYan_PNAS_2017,Bertoldi2017flexible}.

To realize a design strategy that implements directed
aging, we could exploit the strong temperature dependence of
the aging process. For example, we could strongly enhance the rate
at which new functionality is acquired by raising the temperature
towards the glass transition temperature. Lowering the temperature
could then freeze in this functionality.

This approach provides an avenue to reach the “holy grail” of
metamaterials -- producing them on macroscopic scales. Scaling up
metamaterials is usually difficult; if they are designed on a computer,
increasing their size increases the required computer resources.
Moreover, scaling down the “bond size” is perhaps even more difficult
since it requires a precise control of the detailed structure on small
length scales. Directing aging provides a way to alter properties of specific individual bonds by applying only macroscopic strains.

\section*{Acknowledgements}
We are grateful to Heinrich Jaeger, Arvind Murugan and Juan de Pablo for enlightening discussions.  Work was supported by the NSF MRSEC Program DMR-1420709 (NP), NSF DMR- 1404841(SRN) and DOE DE-FG02-03ER46088 (DH) and the Simons Foundation for the collaboration “Cracking the Glass Problem” award $\#$348125 at the University of Chicago. We acknowledge the University of Chicago Research Computing Center for support of this work.

\bibliographystyle{unsrt}
\bibliography{bibliography}

\begin{thebibliography}{10}

\bibitem{Hodge_1995_Science}
Ian~M Hodge.
\newblock Physical aging in polymer glasses.
\newblock {\em Science}, 267(5206):1945--1947, 1995.

\bibitem{Struik_1977_Polymer}
LCE Struik.
\newblock Physical aging in plastics and other glassy materials.
\newblock {\em Polymer Engineering \& Science}, 17(3):165--173, 1977.

\bibitem{Hutchinson_1995_review}
John~M Hutchinson.
\newblock Physical aging of polymers.
\newblock {\em Progress in Polymer Science}, 20(4):703--760, 1995.

\bibitem{mitchell2008aging}
James~K Mitchell.
\newblock Aging of sand--a continuing enigma?
\newblock {\em International Conference on Case Histories in Geotechnical
  Engineering}, 8, 2008.

\bibitem{keim2018memory}
Nathan~C Keim, Joseph Paulsen, Zorana Zeravcic, Srikanth Sastry, and Sidney~R
  Nagel.
\newblock Memory formation in matter.
\newblock {\em arXiv preprint arXiv:1810.08587}, 2018.

\bibitem{Phillips}
JC~Phillips.
\newblock Structural principles of amorphous and glassy semiconductors.
\newblock {\em Journal of Non-Crystalline Solids}, 35:1157--1165, 1980.

\bibitem{Thorpe}
MF~Thorpe and EJ~Garboczi.
\newblock Elastic properties of central-force networks with bond-length
  mismatch.
\newblock {\em Physical Review B}, 42(13):8405, 1990.

\bibitem{CGoodrich_PRL_2015}
Carl~P. Goodrich, Andrea~J. Liu, and Sidney~R. Nagel.
\newblock The principle of independent bond-level response: Tuning by pruning
  to exploit disorder for global behavior.
\newblock {\em Phys. Rev. Lett.}, 114:225501, Jun 2015.

\bibitem{DHex_PRE_2017}
Daniel Hexner, Andrea~J. Liu, and Sidney~R. Nagel.
\newblock Linking microscopic and macroscopic response in disordered solids.
\newblock {\em Phys. Rev. E}, 97:063001, Jun 2018.

\bibitem{DHex_Softmatter_2018}
Daniel Hexner, Andrea~J. Liu, and Sidney~R. Nagel.
\newblock Role of local response in manipulating the elastic properties of
  disordered solids by bond removal.
\newblock {\em Soft Matter}, 14:312--318, 2018.

\bibitem{DReid_PNAS_2018}
Daniel~R. Reid, Nidhi Pashine, Justin~M. Wozniak, Heinrich~M. Jaeger, Andrea~J.
  Liu, Sidney~R. Nagel, and Juan~J. de~Pablo.
\newblock Auxetic metamaterials from disordered networks.
\newblock {\em Proceedings of the National Academy of Sciences},
  115(7):E1384--E1390, 2018.

\bibitem{JRocks_PNAS_2017}
Jason~W. Rocks, Nidhi Pashine, Irmgard Bischofberger, Carl~P. Goodrich,
  Andrea~J. Liu, and Sidney~R. Nagel.
\newblock Designing allostery-inspired response in mechanical networks.
\newblock {\em Proceedings of the National Academy of Sciences},
  114(10):2520--2525, 2017.

\bibitem{LYan_PNAS_2017}
Le~Yan, Riccardo Ravasio, Carolina Brito, and Matthieu Wyart.
\newblock Architecture and coevolution of allosteric materials.
\newblock {\em Proceedings of the National Academy of Sciences},
  114(10):2526--2531, 2017.

\bibitem{Tlusty_PRX_2017}
Tsvi Tlusty, Albert Libchaber, and Jean-Pierre Eckmann.
\newblock Physical model of the genotype-to-phenotype map of proteins.
\newblock {\em Phys. Rev. X}, 7:021037, Jun 2017.

\bibitem{Liu_ARCMP_2010}
Andrea~J. Liu and Sidney~R. Nagel.
\newblock The jamming transition and the marginally jammed solid.
\newblock {\em Annual Review of Condensed Matter Physics}, 1(1):347--369, 2010.

\bibitem{alexander}
Shlomo Alexander.
\newblock Amorphous solids: their structure, lattice dynamics and elasticity.
\newblock {\em Physics reports}, 296(2-4):65--236, 1998.

\bibitem{Wyart_network}
Matthieu Wyart.
\newblock On the rigidity of amorphous solids.
\newblock {\em arXiv preprint cond-mat/0512155}, 2005.

\bibitem{RLakes_Science_1987}
Roderic Lakes.
\newblock Foam structures with a negative poisson{\textquoteright}s ratio.
\newblock {\em Science}, 235(4792):1038--1040, 1987.

\bibitem{Gibson_1982}
Lorna~J Gibson, Michael~Farries Ashby, GS~Schajer, and CI~Robertson.
\newblock The mechanics of two-dimensional cellular materials.
\newblock {\em Proceedings of the Royal Society of London. A. Mathematical and
  Physical Sciences}, 382(1782):25--42, 1982.

\bibitem{Lakes_2017}
Roderic~S. Lakes.
\newblock Negative-poisson's-ratio materials: Auxetic solids.
\newblock {\em Annual Review of Materials Research}, 47(1):63--81, 2017.

\bibitem{SMajumdar2018mechanical}
Sayantan Majumdar, Louis~C Foucard, Alex~J Levine, and Margaret~L Gardel.
\newblock Mechanical hysteresis in actin networks.
\newblock {\em Soft matter}, 14(11):2052--2058, 2018.

\bibitem{TMullin_PRL_2007}
T.~Mullin, S.~Deschanel, K.~Bertoldi, and M.~C. Boyce.
\newblock Pattern transformation triggered by deformation.
\newblock {\em Phys. Rev. Lett.}, 99:084301, Aug 2007.

\bibitem{JShim_SoftMatter_2013}
Jongmin Shim, Sicong Shan, Andrej Košmrlj, Sung~H. Kang, Elizabeth~R. Chen,
  James~C. Weaver, and Katia Bertoldi.
\newblock Harnessing instabilities for design of soft reconfigurable
  auxetic/chiral materials.
\newblock {\em Soft Matter}, 9:8198--8202, 2013.

\bibitem{BFlorijn_PRL_2014}
Bastiaan Florijn, Corentin Coulais, and Martin van Hecke.
\newblock Programmable mechanical metamaterials.
\newblock {\em Phys. Rev. Lett.}, 113:175503, Oct 2014.

\bibitem{DPine_Nature_2005}
David~J Pine, Jerry~P Gollub, John~F Brady, and Alexander~M Leshansky.
\newblock Chaos and threshold for irreversibility in sheared suspensions.
\newblock {\em Nature}, 438(7070):997, 2005.

\bibitem{LCorte_NatureP_2008}
Laurent Corte, Paul~M Chaikin, Jerry~P Gollub, and David~J Pine.
\newblock Random organization in periodically driven systems.
\newblock {\em Nature Physics}, 4(5):420, 2008.

\bibitem{NKeim_PRL_2011}
N~C Keim and Sidney~R. Nagel.
\newblock Generic transient memory formation in disordered systems with noise.
\newblock {\em Physical review letters}, 107 1:010603, 2011.

\bibitem{NKiem_PRE_2013}
Nathan~C. Keim, Joseph~D. Paulsen, and Sidney~R. Nagel.
\newblock Multiple transient memories in sheared suspensions: Robustness,
  structure, and routes to plasticity.
\newblock {\em Phys. Rev. E}, 88:032306, Sep 2013.

\bibitem{JPaunlsen_PRL_2014}
Joseph~D. Paulsen, Nathan~C. Keim, and Sidney~R. Nagel.
\newblock Multiple transient memories in experiments on sheared non-brownian
  suspensions.
\newblock {\em Phys. Rev. Lett.}, 113:068301, Aug 2014.

\bibitem{NKiem_Softmatter_2015}
Nathan~C. Keim and Paulo~E. Arratia.
\newblock Role of disorder in finite-amplitude shear of a 2d jammed material.
\newblock {\em Soft Matter}, 11:1539--1546, 2015.

\bibitem{DFiocco_PRL_2014}
Davide Fiocco, Giuseppe Foffi, and Srikanth Sastry.
\newblock Encoding of memory in sheared amorphous solids.
\newblock {\em Phys. Rev. Lett.}, 112:025702, Jan 2014.

\bibitem{HNagamanasa_PRE_2014}
K.~Hima~Nagamanasa, Shreyas Gokhale, A.~K. Sood, and Rajesh Ganapathy.
\newblock Experimental signatures of a nonequilibrium phase transition
  governing the yielding of a soft glass.
\newblock {\em Phys. Rev. E}, 89:062308, Jun 2014.

\bibitem{Sood_arXiv_2018}
Srimayee {Mukherji}, Neelima {Kandula}, A~K {Sood}, and Rajesh {Ganapathy}.
\newblock {Strength of Mechanical Memories is Maximal at the Yield Point of a
  Soft Glass}.
\newblock {\em arXiv e-prints}, page arXiv:1808.07701, August 2018.

\bibitem{MLavrentovich_PRE_2017}
Maxim~O. Lavrentovich, Andrea~J. Liu, and Sidney~R. Nagel.
\newblock Period proliferation in periodic states in cyclically sheared jammed
  solids.
\newblock {\em Phys. Rev. E}, 96:020101, Aug 2017.

\bibitem{royer2015precisely}
John~R Royer and Paul~M Chaikin.
\newblock Precisely cyclic sand: Self-organization of periodically sheared
  frictional grains.
\newblock {\em Proceedings of the National Academy of Sciences}, 112(1):49--53,
  2015.

\bibitem{corey2008reversible}
Micah Lundberg, Kapilanjan Krishan, Ning Xu, Corey~S O’Hern, and Michael
  Dennin.
\newblock Reversible plastic events in amorphous materials.
\newblock {\em Physical Review E}, 77(4):041505, 2008.

\bibitem{Bertoldi2017flexible}
Katia Bertoldi, Vincenzo Vitelli, Johan Christensen, and Martin van Hecke.
\newblock Flexible mechanical metamaterials.
\newblock {\em Nature Reviews Materials}, 2(11):17066, 2017.

\end{thebibliography}

\end{document}